\documentclass{article}
\usepackage{ijcai23}

\usepackage{times}
\usepackage{soul}
\usepackage{url}
\usepackage[hidelinks]{hyperref}
\usepackage[utf8]{inputenc}
\usepackage[small]{caption}
\usepackage{graphicx}
\usepackage{amsmath}
\usepackage{amsthm}
\usepackage{booktabs}
\usepackage{algorithm}
\usepackage{algorithmic}
\usepackage[switch]{lineno}

\usepackage{amsmath,graphicx,setspace,algorithm,algorithmic,multirow,cases,amssymb,subfigure}
\usepackage{color}
\usepackage{threeparttable}
\title{CiT-Net: Convolutional Neural Networks Hand in Hand with Vision Transformers for Medical Image Segmentation}

\author{
Tao Lei$^{1,2}$
\and
Rui Sun$^{1}$\and
Xuan Wang$^3$\and
Yingbo Wang$^{1}$\and
Xi He$^{1}$\and
Asoke Nandi$^4$
\affiliations
$^1$Shaanxi Joint Laboratory of Artificial Intelligence, Shaanxi University of Science and Technology\\
$^2$Department of Geriatric Surgery, First Affiliated Hospital, Xi’an Jiaotong University\\
$^3$Unmanned System Research Institute, Northwestern Polytechnical University\\
$^4$Department of Electronic and Electrical Engineering, Brunel University London
\emails
leitao@sust.edu.cn,
siri0920@163.com,
wangxuan@nwpu.edu.cn,
\{wangyingbo, xihe\}@sust.edu.cn,
Asoke.Nandi@brunel.ac.uk
}

\begin{document}

\maketitle

%%%%%%%%%%%%%%%%% 0
\begin{abstract}
The hybrid architecture of convolutional neural networks (CNNs) and Transformer are very popular for medical image segmentation. However, it suffers from two challenges. First, although a CNNs branch can capture the local image features using vanilla convolution, it cannot achieve adaptive feature learning. Second, although a Transformer branch can capture the global features, it ignores the channel and cross-dimensional self-attention, resulting in a low segmentation accuracy on complex-content images. To address these challenges, we propose a novel hybrid architecture of convolutional neural networks hand in hand with vision Transformers (CiT-Net) for medical image segmentation. Our network has two advantages. First, we design a dynamic deformable convolution and apply it to the CNNs branch, which overcomes the weak feature extraction ability due to fixed-size convolution kernels and the stiff design of sharing kernel parameters among different inputs. Second, we design a shifted-window adaptive complementary attention module and a compact convolutional projection. We apply them to the Transformer branch to learn the cross-dimensional long-term dependency for medical images. Experimental results show that our CiT-Net provides better medical image segmentation results than popular SOTA methods. Besides, our CiT-Net requires lower parameters and less computational costs and does not rely on pre-training. The code is publicly available at https://github.com/SR0920/CiT-Net.
\end{abstract}

%%%%%%%%%%%%%%%%% 1
\section{Introduction}

Medical image segmentation refers to dividing a medical image into several specific regions with unique properties. Medical image segmentation results can not only achieve abnormal detection of human body regions but also be used to guide clinicians. Therefore, accurate medical image segmentation has become a key component of computer-aided diagnosis and treatment, patient condition analysis, image-guided surgery, tissue and organ reconstruction, and treatment planning. Compared with common RGB images, medical images usually suffer from the problems such as high density noise, low contrast, and blurred edges. So how to quickly and accurately segment specific human organs and lesions from medical images has always been a huge challenge in the field of smart medicine.

In recent years, with the rapid development of computer hardware resources, researchers have continuously developed many new automatic medical image segmentation algorithms based on a large number of experiments. The existing medical image segmentation algorithms can be divided into two categories: based on convolutional neural networks (CNNs) and based on the Transformer networks.

The early traditional medical image segmentation algorithms are based on manual features designed by medical experts using professional knowledge~\cite{suetens2017fundamentals}. These methods have a strong mathematical basis and theoretical support, but these algorithms have poor generalization for different organs or lesions of the human body. Later, inspired by the full convolutional networks (FCN)~\cite{long2015fully} and the encoder-decoder, Ronnebreger et al. designed the U-Net~\cite{ronneberger2015u} network that was first applied to medical image segmentation. After the network was proposed, its symmetric U-shaped encoder and decoder structure received widespread attention. At the same time, due to the small number of parameters and the good segmentation effect of the U-Net network, deep learning has made a breakthrough in medical image segmentation. Then a series of improved medical image segmentation networks are inspired based on the U-Net network, such as 2D U-Net++~\cite{zhou2018unet++}, ResDO-UNet~\cite{liu2023resdo}, SGU-Net~\cite{lei2023sgu}, 2.5D RIU-Net~\cite{lv20222}, 3D Unet~\cite{cciccek20163d}, V-Net~\cite{milletari2016v}, etc. Among them, Alom et al. designed R2U-Net~\cite{alom2018recurrent} by combining U-Net, ResNet~\cite{song2020real}, and recurrent neural network (RCNN)~\cite{girshick2014rich}. Then Gu et al. introduced dynamic convolution~\cite{chen2020dynamic} into U-Net proposed CA-Net~\cite{gu2020net}. Based on U-Net, Yang et al. proposed DCU-Net~\cite{yang2022dcu} by referring to the idea of residual connection and deformable convolution~\cite{dai2017deformable}.Lei et al.~\cite{lei2022semi} proposed a network ASE-Net based on adversarial consistency learning and dynamic convolution.

The rapid development of CNNs in the field of medical image segmentation is largely due to the scale invariance and inductive bias of convolution operation. Although this fixed receptive field improves the computational efficiency of CNNs, it limits its ability to capture the relationship between distant pixels in medical images and lacks the ability to model medical images in a long range.

Aiming at the shortcomings of CNNs in obtaining global features of medical images, scholars have proposed a Transformer architecture. In 2017, Vaswani et al.~\cite{vaswani2017attention} proposed the first Transformer network. Because of its unique structure, Transformer obtains the ability to process indefinite-length input, establish long-range dependency modeling, and capture global information. With the excellent performance of Transformer in NLP fields, ViT~\cite{dosovitskiy2020image} applied Transformer to the field of image processing for the first time. Then Chen et al. put forward TransUNet~\cite{chen2021transunet}, which initiates a new period of Transformer in the field of medical image segmentation. Valanarasu et al. proposed MedT~\cite{valanarasu2021medical} in combination with the gating mechanism. Cao et al. proposed a pure Transformer network Swin-Unet~\cite{cao2021swin} for medical image segmentation, in combination with the shifted-window multi-head self-attention (SW-MSA) in Swin Transformer~\cite{liu2021swin}. Subsequently, Wang et al. designed the BAT~\cite{wang2021boundary} network for dermoscopic images segmentation by combining the edge detection idea~\cite{sun2022survey}. Hatamizadeh et al. proposed Swin UNETR~\cite{tang2022self} network for 3D brain tumor segmentation. Wang et al. proposed the UCTransNet~\cite{wang2022uctransnet} network that combines the channel attention with Transformer.

These methods can be roughly divided into based on the pure Transformer architecture and based on the hybrid architecture of CNNs and Transformer. The pure Transformer network realizes the long-range dependency modeling based on self-attention. However, due to the lack of inductive bias of the Transformer itself, Transformer cannot be widely used in small-scale datasets like medical images~\cite{shamshad2022transformers}. At the same time, Transformer architecture is prone to ignore detailed local features, which reduces the separability between the background and the foreground of small lesions or objects with large-scale changes in the medical image.

The hybrid architecture of CNNs and Transformer realizes the local and global information modeling of medical images by taking advantage of the complementary advantages of CNNs and Transformer, thus achieving a better medical image segmentation effect~\cite{azad2022medical}. However, this hybrid architecture still suffers from the following two problems. First, it ignores the problems of organ deformation and lesion irregularities when modeling local features, resulting in weak local feature expression. Second, it ignores the correlation between the feature map space and the channels when modeling the global feature, resulting in inadequate expression of self-attention. To address the above problems, our main contributions are as follows:

\begin{itemize}
  \item A novel dynamic deformable convolution (DDConv) is proposed. Through task adaptive learning, DDConv can flexibly change the weight coefficient and deformation offset of convolution itself. DDConv can overcome the problems of fixation of receptive fields and sharing of convolution kernel parameters, which are common problems of vanilla convolution and its variant convolutions, such as Atrous convolution and Involution, etc. Improves the ability to perceive tiny lesions and targets with large-scale changes in medical images.
  \item A new (shifted)-window adaptive complementary attention module ((S)W-ACAM) is proposed. (S)W-ACAM realizes the cross-dimensional global modeling of medical images through four parallel branches of weight coefficient adaptive learning. Compared with the current popular attention mechanisms, such as CBAM and Non-Local, (S)W-ACAM fully makes up for the deficiency of the conventional attention mechanism in modeling the cross-dimensional relationship between spatial and channels. It can capture the cross-dimensional long-distance correlation features in medical images, and enhance the separability between the segmented object and the background in medical images.
  \item A new parallel network structure based on dynamically adaptive CNNs and cross-dimensional feature fusion Transformer is proposed for medical image segmentation, called CiT-Net. Compared with the current popular hybrid architecture of CNNs and Transformer, CiT-Net can maximize the retention of local and global features in medical images. It is worth noting that CiT-Net not only abandons pre-training but also has fewer parameters and less computational costs, which are 11.58 M and 4.53 GFLOPs respectively.
\end{itemize}

Compared with the previous vanilla convolution~\cite{ronneberger2015u}, dynamic convolution~\cite{chen2020dynamic}~\cite{li2021involution}, and deformable convolution~\cite{dai2017deformable}, our DDConv can not only adaptively change the weight coefficient and deformation offset of the convolution according to the medical image task, but also better adapt to the shape of organs and small lesions with large-scale changes in the medical image, and additionally, it can improve the local feature expression ability of the segmentation network. Compared with the self-attention mechanism in the existing Transformer architectures~\cite{cao2021swin}~\cite{wang2021boundary}, our (S)W-ACAM requires fewer parameters and less computational costs while it’s capable of capturing the global cross-dimensional long-range dependency in the medical image, and improving the global feature expression ability of the segmentation network. Our CiT-Net does not require a large number of labeled data for pre-training, but it can maximize the retention of local details and global semantic information in medical images. It has achieved the best segmentation performance on both dermoscopic images and liver datasets.

%%%%%%%%%%%%%%%%% 2
\section{Method}

%%%%%%%%%%%%%%%%% 2.1
\subsection{Overall Architecture}

\begin{figure*}[htbp]
	\centerline{\includegraphics[width=\textwidth]{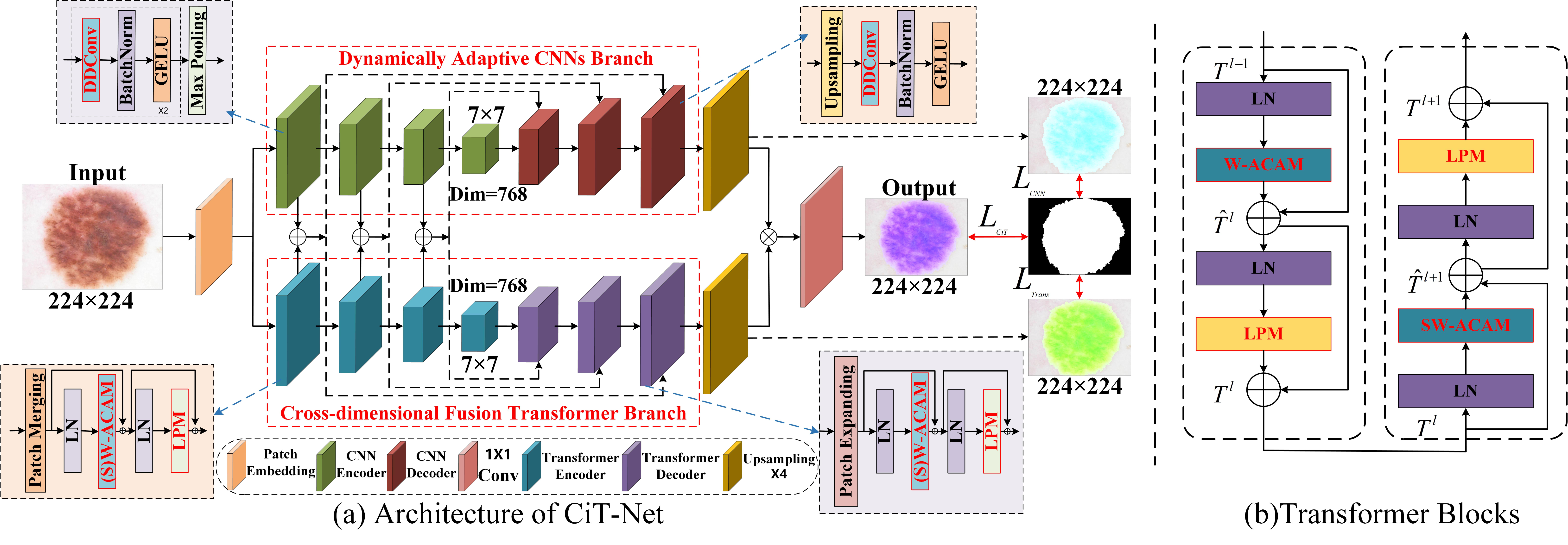}}
	\caption{(a) The architecture of CiT-Net. CiT-Net consists of a dual-branch interaction between dynamically adaptive CNNs and cross-dimensional feature fusion Transformer. The DDConv in the CNNs branch can adaptively change the weight coefficient and deformation offset of the convolution itself, which improves the segmentation accuracy of irregular objects in medical images. The (S)W-ACAM in the Transformer branch can capture the cross-dimensional long-range dependency in medical images, improving the separability of segmented objects and backgrounds in medical images. The lightweight perceptron module (LPM) greatly reduces the parameters and calculations of the original Transformer network by using the Ghost strategy. (b) Two successive Transformer blocks. W-ACAM and SW-ACAM are cross-dimensional self-attention modules with shifted windows and compact convolutional projection configurations.}
	\label{fig2}
\end{figure*}

The fusion of local and global features are clearly helpful for improving medical image segmentation. CNNs capture local features in medical images through convolution operation and hierarchical feature representation. In contrast, the Transformer network realizes the extraction of global features in medical images through the cascaded self-attention mechanism and the matrix operation with context interaction. In order to make full use of local details and global semantic features in medical images, we design a parallel interactive network architecture CiT-Net. The overall architecture of the network is shown in Figure 1 (a). CiT-Net fully considers the complementary properties of CNNs and Transformer. During the forward propagation process, CiT-Net continuously feeds the local details extracted by the CNNs to the decoder of the Transformer branch. Similarly, CiT-Net also feeds the global long-range relationship captured by the Transformer branch to the decoder of the CNNs branch. Obviously, the proposed CiT-Net provides better local and global feature representation than pure CNNs or Transformer networks, and it shows great potential in the field of medical image segmentation.

Specifically, CiT-Net consists of a patch embedding model, dynamically adaptive CNNs branch, cross-dimensional fusion Transformer branch, and feature fusion module. Among them, the dynamically adaptive CNNs branch and the cross-dimensional fusion Transformer branch follow the design of U-Net and Swin-Unet, respectively. The dynamically adaptive CNNs branch consists of seven main stages. By using the weight coefficient and deformation offset adaptive DDConv in each stage, the segmentation network can better understand the local semantic features of medical images, better perceive the subtle changes of human organs or lesions, and improve the ability of extracting multi-scale change targets in medical images. Similarly, the cross-dimensional fusion Transformer branch also consists of seven main stages. By using (S)W-ACAM attention in each stage, as shown in Figure 1 (b), the segmentation network can better understand the global dependency of medical images to capture the position information between different organs, and improve the separability of the segmented object and the background in the medical images.

\begin{figure*}[htbp]
	\centerline{\includegraphics[width=0.8\textwidth]{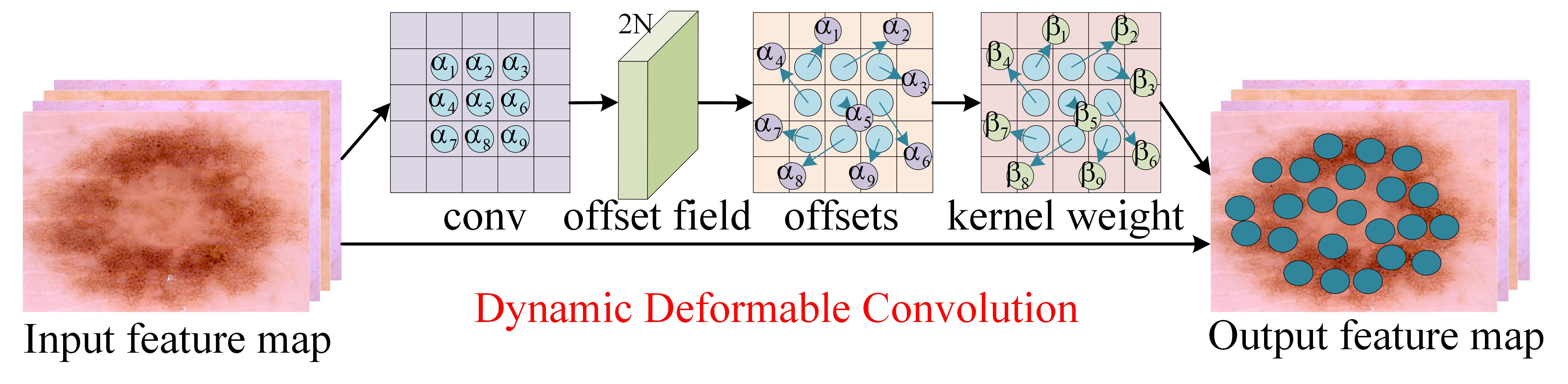}}
	\caption{ The module of the proposed DDConv. Compared with the current popular convolution strategy, DDConv can dynamically adjust the weight coefficient and deformation offset of the convolution itself during the training process, which is conducive to the feature capture and extraction of irregular targets in medical images. $\alpha$ and $\beta$ represent the different weight values of DDConv in different states.}
	\label{fig3}
\end{figure*}

Although our CiT-Net can effectively improve the feature representation of medical images, it requires a large number of training data and network parameters due to the dual-branch structure. As the conventional Transformer network contains a lot of MLP layers, which not only aggravates the training burden of the network but also makes the number of model parameters rise sharply, resulting in the slow training of the model. Inspired by the idea of the Ghost network~\cite{han2020ghostnet}, we redesign the MLP layer in the original Transformer and proposed a lightweight perceptron module (LPM). The LPM can help our CiT-Net not only achieve better medical image segmentation results than MLP but also greatly reduced the parameters and computational complexity of the original Transformer block, even the Transformer can achieve good results without a lot of labeled data training. It is worth mentioning that the dual-branch structure involves mutually symmetric encoders and decoders so that the parallel interaction network structure can maximize the preservation of local features and global features in medical images.
%%%%%%%%%%%%%%%%% 2.2
\subsection{Dynamic Deformable Convolution}

Vanilla convolution has spatial invariance and channel specificity, so it has a limited ability to change different visual modalities when dealing with different spatial locations. At the same time, due to the limitations of the receptive field, it is difficult for vanilla convolution to extract features of small targets or targets with blurred edges. Therefore, vanilla convolution inevitably has poor adaptability and weak generalization ability for complex medical images. Although the existing deformable convolution~\cite{dai2017deformable} and dynamic convolution~\cite{chen2020dynamic}~\cite{li2021involution} outperforms vanilla convolution to a certain extent, they still have the unsatisfied ability to balance the performance and size of networks when dealing with medical image segmentation.

In order to solve the shortcomings of current convolution operations, this paper proposes a new convolution strategy, DDConv, as shown in Figure 2. It can be seen that DDConv can adaptively learn the kernel deformation offset and weight coefficients according to the specific task and data distribution, so as to realize the change of both the shapes and the values of convolution kernels. It can effectively deal with the problems of large data distribution differences and large target deformation in medical image segmentation. Also, DDConv is plug-and-play and can be embedded in any network structure.

The shape change of the convolutional kernel in DDConv is based on the network learning of the deformation offsets. The segmentation network first samples the input feature map $X$ using a square convolutional kernel $S$, and then performs a weighted sum with a weight matrix $M$. The square convolution kernel $S$ determines the range of the receptive field, e.g., a $3 \times 3$ convolution kernel can be expressed as:
\begin{equation}
S=\{( 0,0) ,( 0,1) ,( 0,2) ,...,( 2,1) ,( 2,2)\},
\end{equation}
then the output feature map $Y$ at the coordinate $\varphi_{n}$ can be expressed as:
\begin{equation}
Y\left( \varphi_{n} \right) = {\sum\limits_{\varphi_{m \in S}}{S\left( \varphi_{m} \right)}} \cdot X\left( \varphi_{n} + \varphi_{m} \right),
\end{equation}
when the deformation offset $\bigtriangleup \varphi_{m} = \left\{ m = 1,2,3,\ldots,N \right\}$ is introduced in the weight matrix $M$, $N$ is the total length of $S$. Thus the Equation (2) can be expressed as:
\begin{equation}
Y\left( \varphi_{n} \right) = {\sum\limits_{\varphi_{m \in S}}{S\left( \varphi_{m} \right)}} \cdot X\left( \varphi_{n} + \varphi_{m} + \bigtriangleup \varphi_{m} \right).
\end{equation}

Through network learning, an offset matrix with the same size as the input feature map can be finally obtained, and the matrix dimension is twice that of the input feature map.

To show the convolution kernel of DDConv is dynamic, we first present the output feature map of vanilla convolution:
\begin{equation}
y = \sigma(W \cdot x),
\end{equation}
where $\sigma$ is the activation function, $W$ is the convolutional kernel weight matrix and $y$ is the output feature map. In contrast, the output of the feature map of DDConv is:
\begin{equation}
\hat{y} = \sigma\left( \left( \alpha_{1} \cdot W_{1} + \ldots + \alpha_{n} \cdot W_{n} \right) \cdot x \right),
\end{equation}
where $n$ is the number of weight coefficients, $\alpha_{n}$ is the weight coefficients with learnable parameters and $\hat{y}$ is the output feature map generated by DDConv. DDConv achieves dynamic adjustment of the convolution kernel weights by linearly combining different weight matrices according to the corresponding weight coefficients before performing the convolution operation.

According to the above analysis, we can see that DDConv realizes the dynamic adjustment of the shape and weights of the convolution kernel by combining the convolution kernel deformation offset and the convolution kernel weight coefficient with a minimal number of calculation. Compared with directly increasing the number and size of convolution kernels, the DDConv is simpler and more efficient. The proposed DDConv not only solves the problem of poor adaptive feature extraction ability of fixed-size convolution kernels but also overcomes the defect that different inputs share the same convolution kernel parameters. Consequently, our DDConv can be used to improve the segmentation accuracy of small targets and large targets with blurred edges in medical images.

%%%%%%%%%%%%%%%%% 2.3
\subsection{Shifted Window Adaptive Complementary Attention Module}

\begin{figure}[htbp]
	\centerline{\includegraphics[width=\columnwidth]{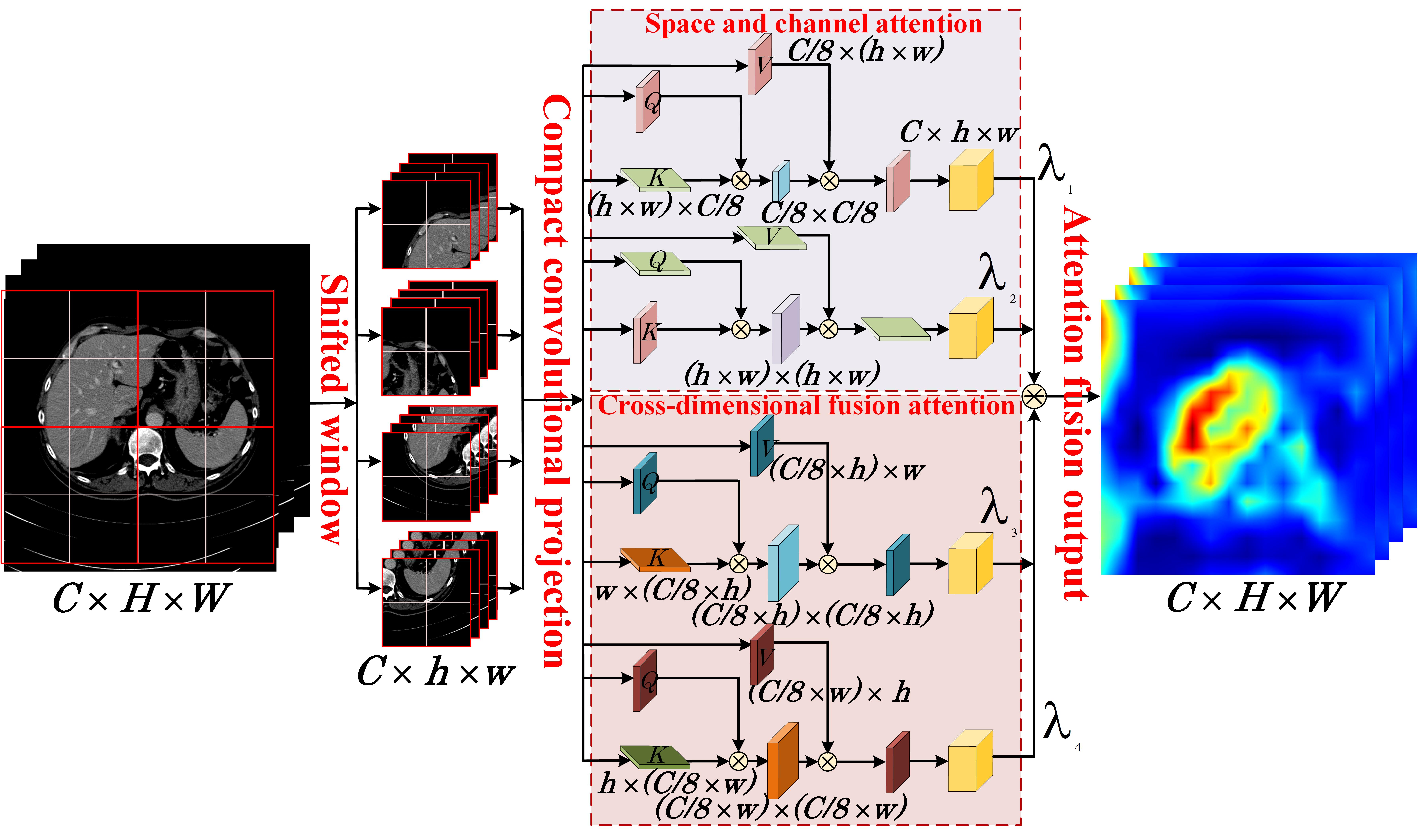}}
	\caption{The module of the proposed (S)W-ACAM. Unlike conventional self-attention, (S)W-ACAM has the advantages of spatial and channel attention, and can also capture long-distance correlation features between spatial and channels. Through the shifted window operation, the spatial resolution of the image is significantly reduced, and through the compact convolutional projection operation, the channel dimension of the image is also significantly reduced. Thus, the overall computational costs and complexity of the network are reduced. $\lambda_{1}$, $\lambda_{2}$, $\lambda_{3}$ and $\lambda_{4}$ are learnable weight parameters.}
	\label{fig4}
\end{figure}

The self-attention mechanism is the core computing unit in Transformer networks, which realizes the capture of long-range dependency of feature maps by utilizing matrix operations. However, the self-attention mechanism only considers the dependency in the spatial dimension but not the cross-dimensional dependency between spatial and channels~\cite{hong2021qau}. Therefore, when dealing with medical image segmentation with low contrast and high density noise, the self-attention mechanism is easy to confuse the segmentation targets with the background, resulting in poor segmentation results.

To solve the problems mentioned above, we propose a new cross-dimensional self-attention module called (S)W-ACAM. As shown in Figure 3, (S)W-ACAM has four parallel branches, the top two branches are the conventional dual attention module~\cite{liu2021scsa} and the bottom two branches are cross-dimensional attention modules. Compared to popular self-attention modules such as spatial self-attention, channel self-attention, and dual self-attention, our proposed (S)W-ACAM can not only fully extract the long-range dependency of spatial and channels, but also capture the cross-dimensional long-range dependency between spatial and channels. These four branches complement each other, provide richer long-range dependency relationships, enhance the separability between the foreground and background, and thus improve the segmentation results for medical images.

The standard Transformer architecture~\cite{dosovitskiy2020image} uses the global self-attention method to calculate the relationship between one token and all other tokens. This calculation method is complex, especially in the face of high-resolution and intensive prediction tasks like medical images where the computational costs will increase exponentially. In order to improve the calculation efficiency, we use the shifted window calculation method similar to that in Swin Transformer~\cite{liu2021swin}, which only calculates the self-attention in the local window. However, in the face of our (S)W-ACAM four branches module, using the shifted window method to calculate self-attention does not reduce the overall computational complexity of the module. Therefore, we also designed the compact convolutional projection. First, we reduce the local size of the medical image through the shifted window operation, then we compress the channel dimension of feature maps through the compact convolutional projection, and finally calculate the self-attention. It is worth mentioning that this method can not only better capture the global high-dimensional information of medical images but also significantly reduce the computational costs of the module. Suppose an image contains $h \times w$ windows, each window size is $M \times M$, then the complexity of the (S)W-ACAM, the global MSA in the original Transformer, and the (S)W-MSA in the Swin Transformer are compared as follows:
\begin{equation}
\Omega\left( {MSA} \right) = 4hwC^{2} + 2(hw)^{2}C,
\end{equation}
\begin{equation}
\Omega\left( {(S)W\mbox{-}MSA} \right) = 4hwC^{2} + 2M^{2}hwC,
\end{equation}
\begin{equation}
\Omega\left( {(S)W\mbox{-}ACAM} \right) = \frac{hwC^{2}}{4} + M^{2}hwC.
\end{equation}
if the former term of each formula is a quadratic function of the number of patches $hw$, the latter term is linear when $M$ is fixed (the default is 7). Then the computational costs of (S)W\mbox{-}ACAM are smaller compared with MSA and (S)W-MSA.

%%% Table 1
\begin{table*}[htbp]
\small
\renewcommand\arraystretch{1.05}
\tabcolsep=0.29cm
\begin{threeparttable}
\begin{tabular}{ccrrrrrrr}
\toprule
\multicolumn{2}{c}{\textbf{Method}}                                                & \textbf{DI}$\uparrow$ & \textbf{JA}$\uparrow$ & \textbf{SE}$\uparrow$ & \textbf{AC}$\uparrow$ & \textbf{SP}$\uparrow$ & \textbf{Para. (M) }$\downarrow$ & \textbf{GFLOPs} \\ \hline
\multirow{5}{*}{\textbf{CNNs}}                            & U-Net~\cite{ronneberger2015u}           & 86.54         & 79.31          & 88.56          & 93.16          & 96.44           & 34.52                & 65.39           \\
                                                          & R2UNet~\cite{alom2018recurrent}        & 87.92         & 80.28          & 90.92          & 93.38          & 96.33           & 39.09                & 152.82          \\
                                                          & Attention Unet~\cite{oktay2018attention} & 87.16         & 79.55          & 88.52          & 93.17          & 95.62           & 34.88                & 66.57           \\
                                                          & CENet~\cite{gu2019net}          & 87.61         & 81.18          & 90.71          & 94.03          & 96.35           & 29.02                & 11.79           \\
                                                          & CPFNet †~\cite{feng2020cpfnet}       & 90.18         & 82.92          & 91.66          & 94.68          & 96.63           & 30.65                & \textcolor{blue}{\textbf{9.15}}            \\ \hline
\multicolumn{1}{l}{\multirow{6}{*}{\textbf{Transformer}}} & Swin-Unet †~\cite{cao2021swin}     & 89.26         & 80.47          & 90.36          & 94.45          & 96.51           & 41.40                & 11.63           \\
\multicolumn{1}{l}{}                                      & TransUNet †~\cite{chen2021transunet}    & 89.39         & 82.10          & 91.43          & 93.67          & 96.54           & 105.30               & 15.21           \\
\multicolumn{1}{l}{}                                      & BAT †~\cite{wang2021boundary}          & 90.21         & 83.49          & 91.59          & 94.85          & 96.57           & 45.56                & 13.38           \\
\multicolumn{1}{l}{}                                      & CvT †~\cite{wu2021cvt}          & 88.23         & 80.21          & 87.60          & 93.68          & 96.28           & 21.51                & 20.53           \\
\multicolumn{1}{l}{}                                      & PVT~\cite{wang2021pyramid}            & 87.31         & 79.99          & 87.74          & 93.10          & 96.21           & 28.86                & 14.92           \\
\multicolumn{1}{l}{}                                      & CrossForm~\cite{wang2021crossformer}      & 87.44         & 80.06          & 88.25          & 93.39          & 96.40           & 38.66                & 13.57           \\ \hline
\multicolumn{2}{c}{\textbf{CiT-Net-T (our)}}                                       & \textcolor{blue}{\textbf{90.72}}         & \textcolor{blue}{\textbf{84.59}}          & \textcolor{blue}{\textbf{92.54}}          & \textcolor{blue}{\textbf{95.21}}          & \textcolor{blue}{\textbf{96.83}}           & \textcolor{red}{\textbf{11.58}}                & \textcolor{red}{\textbf{4.53}}            \\
\multicolumn{2}{c}{\textbf{CiT-Net-B (our)}}                                       & \textcolor{red}{\textbf{91.23}}         & \textcolor{red}{\textbf{84.76}}          & \textcolor{red}{\textbf{92.68}}          & \textcolor{red}{\textbf{95.56}}          & \textcolor{red}{\textbf{98.21}}           & \textcolor{blue}{\textbf{21.24}}                & 13.29          \\ \bottomrule
\end{tabular}
\caption{Performance comparison of the proposed method against the SOTA approaches on the ISIC2018 benchmarks. \textcolor{red}{\textbf{Red}} indicates the best result, and \textcolor{blue}{\textbf{blue}} displays the second-best.}
\begin{tablenotes}
\footnotesize
\item† indicates the model is initialized with pre-trained weights on the ImageNet21K. “Para.” refers to the number of parameters. “GFLOPs” is calculated under the input scale of $224 \times 224$. Since the dermoscopic images are 2D medical images, the comparison methods are all 2D networks.
\end{tablenotes}
\end{threeparttable}
\end{table*}

Among the four parallel branches of (S)W-ACAM, two branches are used to capture channel correlation and spatial correlation, respectively, and the remaining two branches are used to capture the correlation between channel dimension $C$ and space dimension $H$ and vice versa (between channel dimension $C$ and space dimension $W$). After adopting the shifted window partitioning method, as shown in Figure 2 (b), the calculation process of continuous Transformer blocks is as follows:
\begin{equation}
{\hat{T}}^{l} = W\mbox{-}ACAM\left( {LN\left( T^{l - 1} \right)} \right) + T^{l - 1},
\end{equation}
\begin{equation}
T^{l} = LPM\left( {LN\left( {\hat{T}}^{l} \right)} \right) + {\hat{T}}^{l},
\end{equation}
\begin{equation}
{\hat{T}}^{l + 1} = SW\mbox{-}ACAM\left( {LN\left( T^{l} \right)} \right) + T^{l},
\end{equation}
\begin{equation}
T^{l + 1} = LPM\left( {LN\left( {\hat{T}}^{l + 1} \right)} \right) + {\hat{T}}^{l + 1}.
\end{equation}
where ${\hat{T}}^{l}$ and $T^{l}$ represent the output features of (S)W-ACAM and LPM, respectively. W-ACAM represents window adaptive complementary attention, SW-ACAM represents shifted window adaptive complementary attention, and LPM represents lightweight perceptron module. For the specific attention calculation process of each branch, we follow the same principle in Swin Transformer as follows:
\begin{equation}
Attention\left( {Q,K,V} \right) = SoftMax\left( {\frac{QK^{T}}{\sqrt{C/8}} + B} \right)V,
\end{equation}
where relative position bias $B\in {{\mathbb{R}}^{{{M}^{2}}\times {{M}^{2}}}}$, $Q,K,V \in \mathbb{R}^{M^{2} \times \frac{C}{8}}$ are query, key, and value matrices respectively. $\frac{C}{8}$ represents the dimension of query/key, and $M^{2}$ represents the number of patches. 

After four parallel attention branches ${Out}_{1}$, ${Out}_{2}$, ${Out}_{3}$ and ${Out}_{4}$ are calculated, the final feature fusion output is:
\begin{equation}
Out = \lambda_{1} \cdot {Out}_{1} + \lambda_{2} \cdot {Out}_{2} + \lambda_{3} \cdot {Out}_{3} + \lambda_{4} \cdot {Out}_{4},
\end{equation}
where $\lambda_{1}$, $\lambda_{2}$, $\lambda_{3}$ and $\lambda_{4}$ are learnable parameters that enable adaptive control of the importance of each attention branch for spatial and channel information in a particular segmentation task through the back-propagation process of the segmentation network.

Different from other self-attention mechanisms, the (S)W-ACAM in this paper can fully capture the correlation between spatial and channels, and reasonably use the context information of medical images to achieve long-range dependence modeling. Since our (S)W-ACAM effectively overcomes better feature representation of the defect that the conventional self-attention only focuses on the spatial self-attention of images and ignores the channel and cross-dimensional self-attention, it achieves the best image suffers from large noise, low contrast, and complex background.

%%%%%%%%%%%%%%%%% 2.4
\subsection{Architecture Variants}
We have built a CiT-Net-T as a base network with a model size of 11.58 M and a computing capacity of 4.53 GFLOPs. In addition, we built the CiT-Net-B network to make a fair comparison with the latest networks such as CvT~\cite{wu2021cvt} and PVT~\cite{wang2021pyramid}. The window size is set to 7, and the input image size is $224 \times 224$. Other network parameters are set as follows:

\begin{itemize}
  \item CiT-Net-T: $layer~number = \left\{ 2,~2,~6,~2,~6,~2,~2 \right\}$, $H = \left\{ 3,~6,~12,~24,~12,~6,~3 \right\}$, $D = 96$
  \item CiT-Net-B: $layer~number = \left\{ 2,~2,~18,~2,~18,~2,~2 \right\}$, $H = \left\{ 4,~8,~16,~32,~16,~8,~4 \right\}$, $D = 96$, 
\end{itemize}

$D$ represents the number of image channels when entering the first layer of the dynamically adaptive CNNs branch and the cross-dimensional fusion Transformer branch, $layer ~number$ represents the number of Transformer blocks used in each stage, and $H$ represents the number of multiple heads in self-attention.

%%%%%%%%%%%%%%%%% 3
\section{Experiment and Results}
%%%%%%%%%%%%%%%%% 3.1
\subsection{Datasets}
We conducted experiments on the skin lesion segmentation dataset ISIC2018 from the International Symposium on Biomedical Imaging (ISBI) and the Liver Tumor Segmentation Challenge dataset (LiTS) from the Medical Image Computing and Computer Assisted Intervention Society (MICCAI). The ISIC2018 contains 2,594 dermoscopic images for training, but the ground truth images of the testing set have not been released, thus we performed a five-fold cross-validation on the training set for a fair comparison. The LiTS contains 131 3D CT liver scans, where 100 scans of which are used for training, and the remaining 31 scans are used for testing. In addition, all images are empirically resized to $224 \times 224$ for efficiency.

%%%%%%%%%%%%%%%%% 3.2
\subsection{Implementation Details}
All the networks are implemented on NVIDIA GeForce RTX 3090 24GB and PyTorch 1.7. We utilize Adam with an initial learning rate of 0.001 to optimize the networks. The learning rate decreases in half when the loss on the validation set has not dropped by 10 epochs. We used mean squared error loss (MSE) and Dice loss as loss functions in our experiment.

%%% Table 2
\begin{table*}[htbp]
\small
\renewcommand\arraystretch{1.05}
\tabcolsep=0.1cm
\begin{threeparttable}
\begin{tabular}{lcrrrrrrr}\toprule
\multicolumn{2}{c}{\textbf{Method}}                                         & \textbf{DI} $\uparrow$       & \textbf{VOE} $\downarrow$      & \textbf{RVD} $\downarrow$     & \textbf{ASD} $\downarrow$     & \textbf{RMSD} $\downarrow$      & \textbf{Para. (M)} $\downarrow$ & \textbf{GFLOPs} \\ \hline
\multicolumn{1}{c}{\multirow{6}{*}{\textbf{CNNs}}} & U-Net~\cite{ronneberger2015u}           & 93.99±1.23 & 11.13±2.47 & 3.22±0.20 & 5.79±0.53 & 123.57±6.28 & 34.52       & 65.39  \\
\multicolumn{1}{c}{}                      & R2UNet~\cite{alom2018recurrent}         & 94.01±1.18 & 11.12±2.37 & 2.36±0.15 & 5.23±0.45 & 120.36±5.03 & 39.09       & 152.82 \\
\multicolumn{1}{c}{}                      & Attention Unet~\cite{oktay2018attention} & 94.08±1.21 & 10.95±2.36 & 3.02±0.18 & 4.95±0.48 & 118.67±5.31 & 34.88       & 66.57  \\
\multicolumn{1}{c}{}                      & CENet~\cite{gu2019net}          & 94.04±1.15 & 11.03±2.31 & 6.19±0.16 & 4.11±0.51 & 115.40±5.82 & 29.02       & \textcolor{blue}{\textbf{11.79}}  \\
\multicolumn{1}{c}{}                      & 3D Unet~\cite{cciccek20163d}         & 94.10±1.06 & 11.13±2.23 & \textcolor{blue}{\textbf{1.42±0.13}} & 2.61±0.45 & 36.43±5.38  & 40.32       & 66.45  \\
\multicolumn{1}{c}{}                      & V-Net~\cite{milletari2016v}           & 94.25±1.03 & 10.65±2.17 & 1.92±0.11 & 2.48±0.38 & 38.28±5.05  & 65.17       & 55.35  \\ \hline
\multirow{5}{*}{\textbf{Transformer}}              & Swin-Unet †~\cite{cao2021swin}     & 95.62±1.32 & 9.73±2.16  & 2.78±0.21 & 2.35±0.35 & 38.85±5.42  & 41.40       & 11.63  \\
                                          & TransUNet †~\cite{chen2021transunet}    & 95.79±1.09 & 9.82±2.10  & 1.98±0.15 & 2.33±0.41 & 37.22±5.23  & 105.30      & 15.21  \\
                                          & CvT †~\cite{wu2021cvt}          & 95.81±1.25 & 9.66±2.31  & 1.77±0.16 & 2.34±0.29 & 36.71±5.09  & 21.51       & 20.53  \\
                                          & PVT~\cite{wang2021pyramid}            & 94.56±1.15 & 9.75±2.19  & 1.69±0.12 & 2.42±0.34 & 37.35±5.16  & 28.86       & 14.92  \\
                                          & CrossForm~\cite{wang2021crossformer}      & 94.63±1.24 & 9.72±2.24  & 1.65±0.15 & 2.39±0.31 & 37.21±5.32  & 38.66       & 13.57  \\ \hline
\multicolumn{2}{c}{\textbf{CiT-Net-T (our)}}                                & \textcolor{blue}{\textbf{96.48±1.05}} & \textcolor{blue}{\textbf{9.53±2.11}}  & 1.45±0.12 & \textcolor{blue}{\textbf{2.29±0.33}} & \textcolor{blue}{\textbf{36.21±4.97}}  & \textcolor{red}{\textbf{11.58}}       & \textcolor{red}{\textbf{4.53}}   \\
\multicolumn{2}{c}{\textbf{CiT-Net-B (our)}}                                & \textcolor{red}{\textbf{96.82±1.22}} & \textcolor{red}{\textbf{9.46±2.33}}  & \textcolor{red}{\textbf{1.38±0.13}} & \textcolor{red}{\textbf{2.21±0.35}} & \textcolor{red}{\textbf{36.08±4.88}}  & \textcolor{blue}{\textbf{21.24}}       & 13.29  \\ \bottomrule
\end{tabular}
\caption{Performance comparison of the proposed method against the SOTA approaches on the LiTS-Liver benchmarks. \textcolor{red}{\textbf{Red}} indicates the best result, and \textcolor{blue}{\textbf{blue}} displays the second-best.}
\begin{tablenotes}
\footnotesize
\item† indicates the model initialized with pre-trained weights on ImageNet21K. “Para.” refers to the number of parameters. “GFLOPs” is calculated under the input scale of $224 \times 224$. Compared with the comparison experiment oin the ISIC2018 dataset, 3D Unet and V-Net are introduced into the comparison experiment oin the LiTS-Liver dataset.
\end{tablenotes}
\end{threeparttable}
\end{table*}

%%%%%%%%%%%%%%%%% 3.3
\subsection{Evaluation and Results}
In this paper, we selected the mainstream medical image segmentation networks U-Net~\cite{ronneberger2015u}, Attention Unet~\cite{oktay2018attention}, Swin-Unet~\cite{cao2021swin}, PVT~\cite{wang2021pyramid}, CrossForm~\cite{wang2021crossformer} and the proposed CiT-Net to conduct a comprehensive comparison of the two different modalities datasets, ISIC2018 and the LiTS.

In the experiment of the ISIC2018 dataset, we made an overall evaluation of the mainstream medical image segmentation network by using five indicators: Dice (DI), Jaccard (JA), Sensitivity (SE), Accuracy (AC), and Specificity (SP). Table 1 shows the quantitative analysis of the results of the proposed CiT-Net and the current mainstream CNNs and Transformer networks in the ISIC2018 dataset. From the experimental results, we can conclude that our CiT-Net has the minimum number of parameters and the lowest computational costs, and can obtain the best segmentation effect on the dermoscopic images without adding pre-training. Moreover, our CiT-Net-T network has only 11.58 M of parameters and 4.53 GFLOPs of computational costs, but still achieves the second-best segmentation effect. Our CiT-Net-B network, BAT, CvT, and CrossForm have similar parameters or computational costs, but in the ISIC2018 dataset, the division Dice value of our CiT-Net-B is 1.02\%, 3.00\%, and 3.79\% higher than that of the BAT, CvT, and CrossForm network respectively. In terms of other evaluation indicators, our CiT-Net-B is also significantly better than other comparison methods.

In the experiment of the LiTS-Liver dataset, we conducted an overall evaluation of the mainstream medical image segmentation network by using five indicators: DI, VOE, RVD, ASD and RMSD. Table 2 shows the quantitative analysis of the results of the proposed CiT-Net and the current mainstream networks in the LiTS-Liver dataset. It can be seen from the experimental results that our CiT-Net has great advantages in medical image segmentation, which further verifies the integrity of CiT-Net in preserving local and global features in medical images. It is worth noting that the CiT-Net-B and CiT-Net-T networks have achieved good results in medical image segmentation in the first and second place, with the least number of model parameters and computational costs. The division Dice value of our CiT-Net-B network without pre-training is 1.20\%, 1.03\%, and 1.01\% higher than that of the Swin-Unet, TransUNet, and CvT network with pre-training. In terms of other evaluation indicators, our CiT-Net-B is also significantly better than other comparison methods.

\begin{table}[htbp]
\footnotesize
\renewcommand\arraystretch{0.8}
\tabcolsep=0.015cm
\begin{tabular}{lcccrr}
\toprule
\multicolumn{1}{c}{\textbf{Backbone}} & \multicolumn{1}{c}{\textbf{DDConv}} & \multicolumn{1}{c}{\textbf{(S)W-ACAM}} & \multicolumn{1}{c}{\textbf{LPM}} & \multicolumn{1}{c}{\textbf{Para. (M)}} & \multicolumn{1}{c}{\textbf{DI (\%) $\uparrow$}} \\ \hline
U-Net+Swin-Unet              &                            &                             &                         & 46.92                         & 87.45                         \\
U-Net+Swin-Unet              & $\surd$                          &                             &                         & 48.25                         & 89.15                         \\
U-Net+Swin-Unet              &                            & $\surd$                           &                         & 30.26                         & 89.62                         \\
U-Net+Swin-Unet              &                            &                             & $\surd$                       & 15.45                         & 88.43                         \\
U-Net+Swin-Unet              & $\surd$                          & $\surd$                           &                         & 32.16                         & 90.88                         \\
U-Net+Swin-Unet              & $\surd$                    &                                   & $\surd$                 & 16.93                         & 89.12                        \\
U-Net+Swin-Unet              &                            & $\surd$                           & $\surd$                       & 9.67                          & 89.46                         \\
CiT-Net-T (our)              & $\surd$                          & $\surd$                           & $\surd$                       & 11.58                         & 90.72               \\ \bottomrule  
\end{tabular}
\caption{Ablation experiments of DDConv, (S)W-ACAM and LPM in CiT-Net in the ISIC2018 dataset.}
\end{table}

%%%%%%%%%%%%%%%%% 3.4
\subsection{Ablation Study}

In order to fully prove the effectiveness of different modules in our CiT-Net, we conducted a series of ablation experiments on the ISIC2018 dataset. As shown in Table 3, we can see that the Dynamic Deformable Convolution (DDConv) and (Shifted) Window Adaptive Complementary Attention Module ((S)W-ACAM) proposed in this paper show good performance, and the combination of these two modules, CiT-Net shows the best medical image segmentation effect. At the same time, the Lightweight Perceptron Module (LPM) can significantly reduce the overall parameters of the CiT-Net.

%%%%%%%%%%%%%%%%% 4
\section{Conclusion}
In this study, we have proposed a new architecture CiT-Net that combines dynamically adaptive CNNs and cross-dimensional fusion Transformer in parallel for medical image segmentation. The proposed CiT-Net integrates the advantages of both CNNs and Transformer, and retains the local details and global semantic features of medical images to the maximum extent through local relationship modeling and long-range dependency modeling. The proposed DDConv overcomes the problems of fixed receptive field and parameter sharing in vanilla convolution, enhances the ability to express local features, and realizes adaptive extraction of spatial features. The proposed (S)W-ACAM self-attention mechanism can fully capture the cross-dimensional correlation between feature spatial and channels, and adaptively learn the important information between spatial and channels through network training. In addition, by using the LPM to replace the MLP in the traditional Transformer, our CiT-Net significantly reduces the number of parameters, gets rid of the dependence of the network on pre-training, avoids the challenge of the lack of labeled medical image data and easy over-fitting of the network. Compared with popular CNNs and Transformer medical image segmentation networks, our CiT-Net shows significant advantages in terms of operational efficiency and segmentation effect.

\section*{Acknowledgments}
This work was supported in part by the National Natural Science Foundation of China under Grants 62271296, 62201334 and 62201452, in part by the Natural Science Basic Research Program of Shaanxi under Grant 2021JC-47, and in part by the Key Research and Development Program of Shaanxi under Grants 2022GY-436 and 2021ZDLGY08-07.

%% The file named.bst is a bibliography style file for BibTeX 0.99c
\bibliographystyle{named}
\bibliography{ijcai23}

\end{document}